\documentclass[%
 reprint,
superscriptaddress,
 amsmath,amssymb,
prb,
]{revtex4-2}

\usepackage{amsmath}
\usepackage{graphicx}
\usepackage{dcolumn}
\usepackage{bm}
\usepackage{hyperref}
\hypersetup{hypertex=true,
	colorlinks=true,
	linkcolor=blue,
	anchorcolor=blue,
	urlcolor=blue,
	citecolor=blue}

\begin{document}

\preprint{APS/123-QED}

\title{Strain-modulated Valley Polarization and Piezomagnetic Effects in Altermagnetic Cr$_2$S$_2$}

\author{Chen Chen}
\affiliation{School of Science, Nanjing University of Posts and Telecommunications (NJUPT), Nanjing 210023, China.}

\author{Xiaoyang He}
\affiliation{School of Science, Nanjing University of Posts and Telecommunications (NJUPT), Nanjing 210023, China.}

\author{Qizhen Xiong}
\affiliation{The Faculty of Mathematical \& Physical Sciences, University College London(UCL), London WC1E6BT, England.}

\author{Chuye Quan}
\affiliation{Institute of Advanced Materials (IAM), Nanjing University of Posts and Telecommunications (NJUPT), Nanjing 210023, China.}

\author{Haojie Hou}
\affiliation{School of Science, Nanjing University of Posts and Telecommunications (NJUPT), Nanjing 210023, China.}

\author{Shilei Ji}
\email{njuptjishilei@outlook.com}
\affiliation{School of Science, Nanjing University of Posts and Telecommunications (NJUPT), Nanjing 210023, China.}

\author{Jianping Yang}
\affiliation{School of Science, Nanjing University of Posts and Telecommunications (NJUPT), Nanjing 210023, China.}

\author{Xing'ao Li}
\email{lxahbmy@126.com}
\affiliation{School of Science, Nanjing University of Posts and Telecommunications (NJUPT), Nanjing 210023, China.}
\affiliation{Institute of Advanced Materials (IAM), Nanjing University of Posts and Telecommunications (NJUPT), Nanjing 210023, China.}
\affiliation{College of science, Zhejiang University of Science and Technology, Hangzhou 310023, China.}


\date{\today}

\begin{abstract}
Altermagnetism exhibits advantages over both ferromagnetic and antiferromagnetic counterparts by enabling spin splitting within antiferromagnetic materials. Currently, it is established that valley polarization in altermagnetism remains largely insensitive to spin-orbit coupling and spin. Here, using Cr$_2$S$_2$ as a case study, we investigate the mechanism through which an external field modulates valley polarization in altermagnetism. This effect arises from the external field's disruption of diagonal mirror symmetry $M_{xy}$, consequently inducing valley polarization within the material. Strain not only induces valley polarization but also generates an almost uniform magnetic field, which can reach as high as 118.39 T under 5\% uniaxial strain. In addition, this symmetry breaking in Cr$_2$S$_2$ monolayers results in significant piezomagnetic properties, merging piezomagnetic and altermagnetic characteristics in two-dimensional materials. 
\end{abstract}

\maketitle

\section{\label{sec:level1} INTRODUCTION }
With continuous advancements in science and technology, magnetic materials are playing an increasingly important role in scientific research.\cite{RN737, RN734, RN733} Recently, a new type of magnetism called `altermagnetism' has garnered significant attention as a third category distinct from traditional ferromagnetism (FM) and antiferromagnetism (AFM).\cite{RN736, RN749} Current research in spintronics primarily focuses on exploring transport and quantum states dominated by FM.\cite{RN776,RN712,RN713} However, FM is susceptible to external interference, causing magnetic instability. AFM, characterized by magnetic field stability and ultrafast dynamics\cite{RN741,RN58,RN440}, maintains zero magnetic moment and PT symmetry. Altermagnets have dual properties similar to ferromagnets and antiferromagnets. On one hand, like AFM, altermagnets possess a zero magnetic moment and are resistant to external magnetic field interference. On the other hand, similar to FM, the spin-dependent bands of the altermagnet also split at the K point, and the band splitting is spread throughout the k space, resulting in FM related quantum transport characteristics. This discovery not only advances spintronics research but also provides new insights into potential superconducting states in various magnetic materials.\cite{RN641,RN744} Several two-dimensional materials have been predicted to exhibit altermagnetism, such as RuO$_{2}$\cite{RN738}, MnTe\cite{RN751}, and Cr$_{2}$SO\cite{RN750}. 

The integration of charge and spin degrees of freedom with valley degrees of freedom\cite{RN776, RN592} in ferrovalley materials has emerged as a prominent research focus in valleytronics, holding significant promise for encoding, manipulating, and transmitting information.\cite{RN787,RN746,RN781} Recent experiments have spotlighted two-dimensional transition-metal dichalcogenides (TMDCs)\cite{RN778, RN520, RN598} with strong spin-orbit coupling (SOC) in the realm of valley electronics. In TMDCs, valley polarization is absent owing to time reversal symmetry protection. Various approaches have been explored to induce valley polarization, including optical pumping\cite{RN742, RN597, RN730}, magnetic doping\cite{RN593, RN745}, external magnetic fields\cite{RN752, RN642}, and magnetic proximity\cite{RN508} have been explored. Recently, ferrovalley two-dimensional materials with intrinsic valley polarization have been proposed.\cite{RN564, RN520} Importantly, altermagnetic candidates are particularly appealing for valley electronics applications due to their high storage density, resilience to external magnetic fields, and ultra-fast write speeds.\cite{RN720, RN550, RN560} 

To assess the electromagnetic characteristics of materials, we focus on the piezomagnetic effect. Traditional piezomagnetism refers to the phenomenon wherein the application of an external magnetic field alters the internal magnetization of a magnetic material.\cite{RN305,RN785} Instead of rearranging the magnetic moments inside the material, the piezomagnetism here is caused by a strain-induced occupation imbalance between spin-up and spin-down electrons, which is mainly determined by the anisotropy of electronic properties.\cite{RN686,RN784}

In this paper, we predict Cr$_2$S$_2$ is a 2D altermagnetic material  and provide a detailed analysis of the mechanisms by which an external field modulates valley polarization in such materials, using Cr$_2$S$_2$ as a case study. Cr$_2$S$_2$ monolayer with spatial symmetry group $P4-mmm$ is proposed based on first principles, and its stability, magnetic ground state, band structure and magnetic anisotropy are investigated in detail. Cr$_2$S$_2$ monolayers exhibit AFM semiconductor properties with in-plane magnetic anisotropy character. We observe the generation of an almost uniform effective magnetic field under strain conditions, achieving a strength of 118.39 T when the uniaxial strain is 5\%.  Interestingly, the Cr$_2$S$_2$ monolayer has piezomagnetic properties. The magnetization size and direction can be adjusted by electron doping concentration and uniaxial strain strength. It provides a platform for the application of spin device. It is important to highlight that under biaxial tensile strain, the Cr$_2$S$2$ monolayer fails to achieve valley polarization and piezomagnetic properties due to the inability of biaxial strain to alter the $M_{xy}$ symmetry of the crystal. This is because the biaxial strain does not change the $M_{xy}$ symmetry of the crystal.

\section{COMPUTATIONAL DETAILS}
Within the framework of density functional theory (DFT), we perform spin-polarized first-principles calculations using the standard VASP code and the projector augmented-wave (PAW) method.\cite{RN729,RN735,RN739} We employ the generalized gradient approximation of Perdew-Burke-Ernzerhof (PBE-GGA) as the exchange-correlation functional.\cite{RN14} The kinetic energy cutoff is set to 500 eV, total energy convergence criterion of 10$^{-8}$ eV, and force convergence criterion of 0.0001 eV/Å. To describe Cr 3d electrons, the field coulomb of Cr 3d electrons is set to $U = 2.26$ eV.\cite{RN722,RN636} A vacuum space of more than 20 Å is imposed to avoid interlayer interactions. A 14×14×1 Monkhorst-Pack k-point is employed to sample the first Brillouin zone. The SOC effect is
considered to study the band structure and magnetocrystalline anisotropy of the Cr$_2$S$_2$ monolayer under strain. The Phonon dispersion spectrum of Cr$_2$S$_2$ monolayers  is obtained using the PHONOPY code with a 2×2×1 supercell.\cite{RN418}
\section{RESULTS AND DISCUSSION}

\subsection{\label{sec:citeref}STRAIN-MODULATED VALLEY POLARIZATION}
We analyze the mechanism by which uniaxial strain modulates valley polarisation and explain why biaxial strain fails to induce valley polarisation. The energy gain at the extreme valley (assuming non-degenerate) is\cite{RN686}
\begin{equation}\label{eqn-2}
\Delta E=\sum_{\mathrm{ij}} d_{i j} \varepsilon_{i j},
\end{equation}
where d is the the tensor of deformation potential and $\varepsilon$ is the strain tensor. The $i, j \in x,y$. The diagonal mirror symmetry $M_{xy}$ is expressed as
\begin{equation}\label{eqn-3}
M_{x y}=\left(\begin{array}{cc}
0 & -1 \\
-1 & 0
\end{array}\right).
\end{equation}
Under external strain, the energy displacement of valley X is
\begin{equation}\label{eqn-4}
\Delta E_X=d_{x x} \varepsilon_{x x}+d_{y y} \varepsilon_{y y}+2 d_{x y} \varepsilon_{x y},
\end{equation}
and the energy displacement of valley Y can be calculated as
\begin{equation}\label{eqn-5}
\Delta E_Y=d_{y y} \varepsilon_{x x}+d_{x x} \varepsilon_{y y}+2 d_{x y} \varepsilon_{x y}.
\end{equation}
The energy displacement of the X and Y valleys under uniaxial strain along the X axis is
\begin{equation}\label{eqn-6}
\begin{aligned}
& \Delta E_X=d_{x x} \varepsilon_{x x} \\
& \Delta E_Y=d_{y y} \varepsilon_{x x},
\end{aligned}
\end{equation}
then the relative energy displacement of the two valleys is
\begin{equation}\label{eqn-7}
\delta E^{V / C}=\Delta E_{X}^{{V} /{C}}-\Delta E_{Y}^{{V} /{C}}=\eta^{\mathrm{V} / \mathrm{C}} \varepsilon_{\mathrm{xx}},
\end{equation}
where $\delta \mathrm{E}^{\mathrm{V} / \mathrm{C}}$ is the size of valley splitting of conduction and valence band, $\eta^{\mathrm{V} / \mathrm{C}}=(d_{x x}-d_{y y})$ is a constant. The mirror symmetry $M_{xy}$ is broken due to the presence of uniaxial strain $\varepsilon_{x x}$, which leads to the emergence of valley polarization. According to Equation\eqref{eqn-7}, the relationship between valley splitting size and strain strength is linear. By fitting \hyperref[fgr4]{Fig. \ref{fgr4}(a)}, we get $\eta^{\mathrm{V}}=-18.46$, $\eta^{\mathrm{C}}=12.22$. The energy displacement of X and Y valleys under biaxial strain is

\begin{equation}\label{eqn-8}
\Delta E_X=\Delta E_Y=2 d_{x y} \varepsilon_{x y},
\end{equation}
then the relative energy displacement of the two valleys is

\begin{equation}\label{eqn-9}
\delta E^{V / C}=0,
\end{equation}
this indicates that biaxial strain does not produce valley splitting, consistent with the results in \hyperref[fgr4]{Fig. \ref{fgr4}(b)}.

\subsection{\label{sec:level2}ATOMIC AND MAGNETIC STRUCTURES}

\begin{figure}[t]
	\centering
	\includegraphics[width=\linewidth]{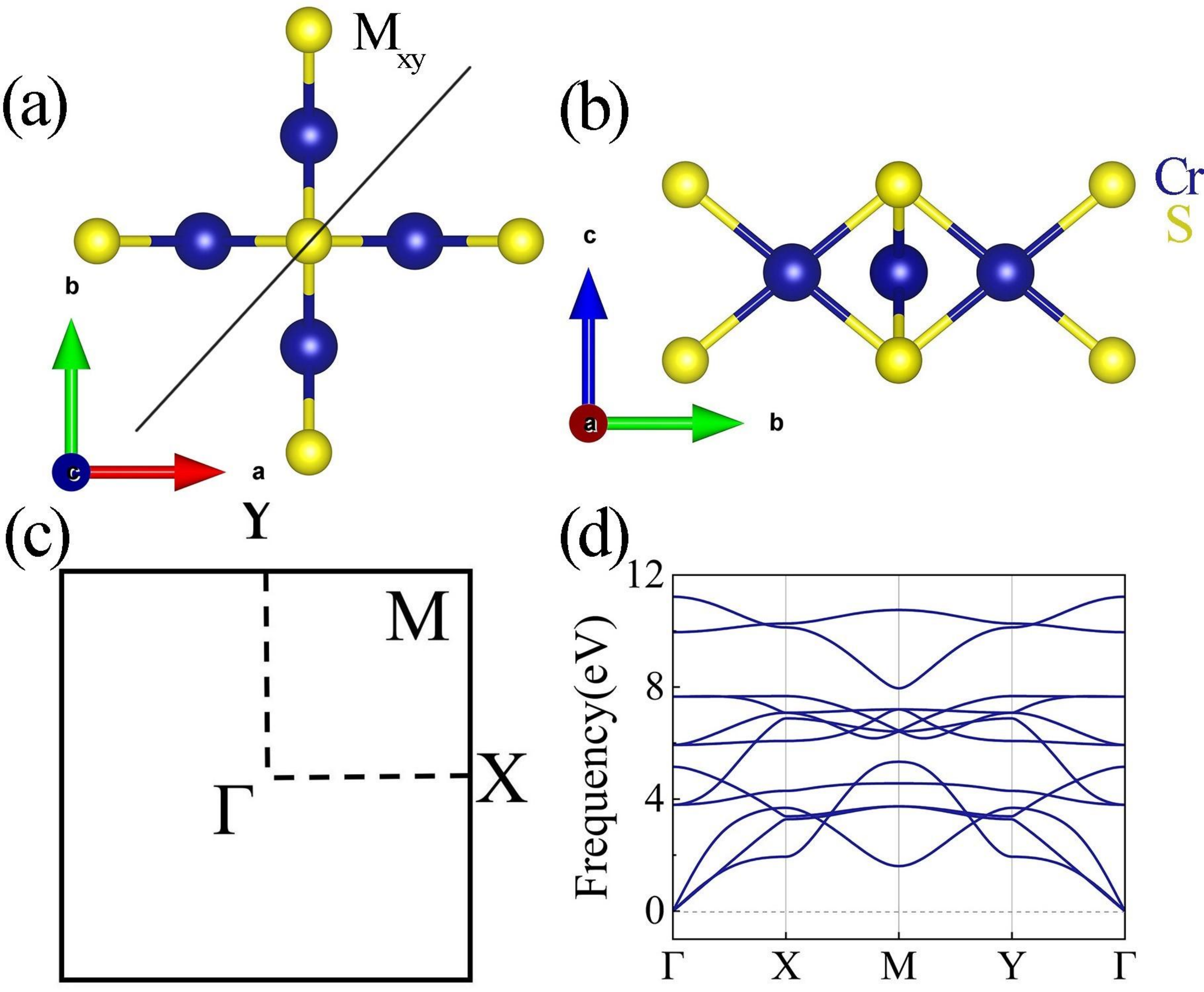}
	\caption{Lattice and phonon spectrum of Cr$_2$S$_2$ monolayer (a) and (b) Top view and side view of Cr$_2$S$_2$ monolayer, where the balls of dark blue and yellow label the Cr and S atoms, respectively. (c)The first Brillouin zone with high symmetry points.(d)Phonon spectrums of monolayer Cr$_2$S$_2$. }
	\label{fgr1}
\end{figure}

\begin{figure}[t]
	\centering
	\includegraphics[width=7.5cm]{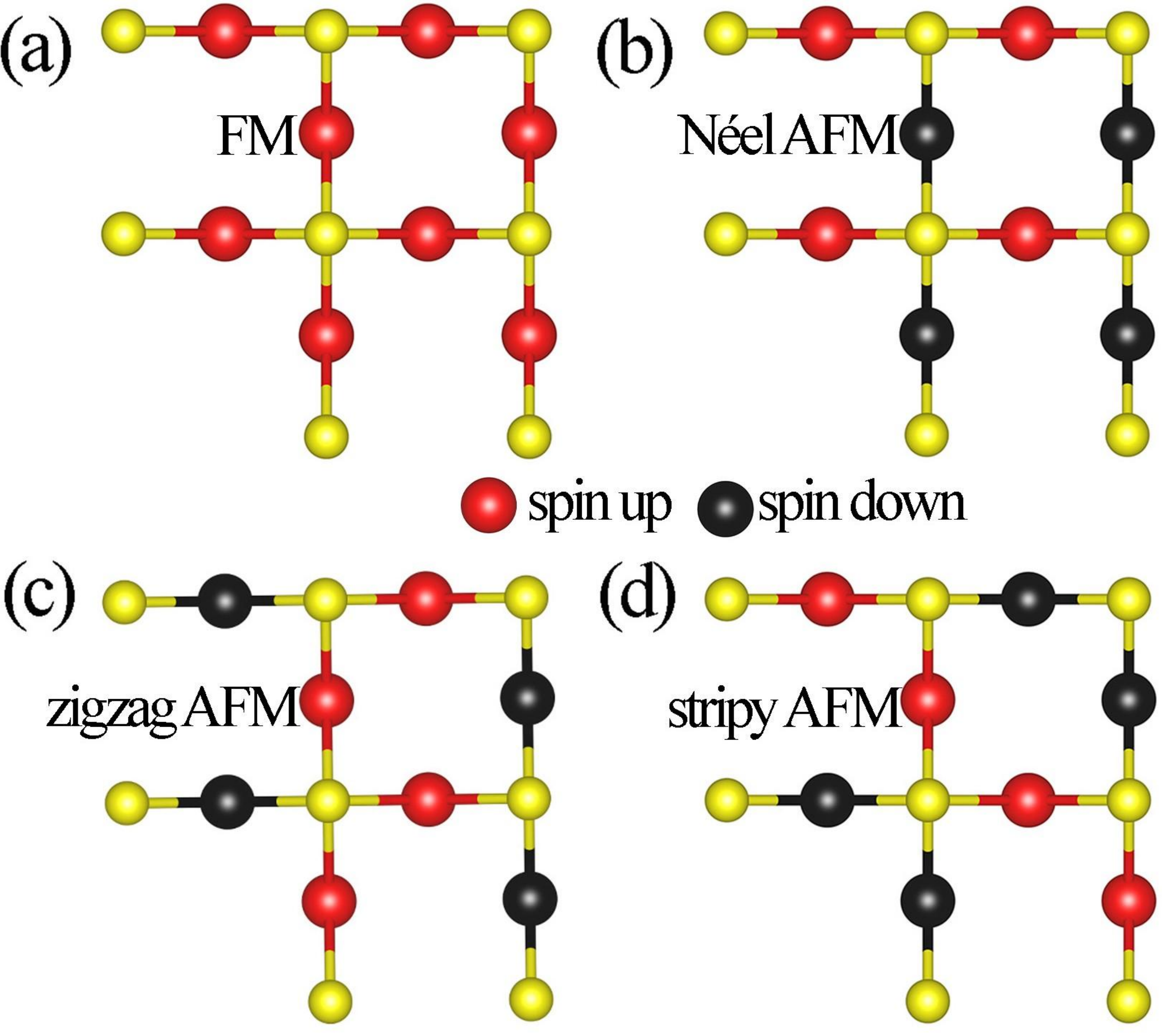}
	\caption{ (a) FM, (b) N\'{e}el AFM, (c) zigzag AFM and (d) stripy AFM configurations of the Cr$_2$S$_2$ monolayer. Red and black balls indicate spin-up and spin-down directions, respectively. 
	}
	\label{fgr2}
\end{figure}

\hyperref[fgr1]{Figs. \ref{fgr1}(a) and (b)} illustrate the crystal structure of a Cr$_2$S$_2$ monolayer, which has three atomic layers, the middle layer is composed of Cr atoms, and the upper and lower layers are S atoms. Each Cr atom is surrounded by four S atoms, and the unit cell of monolayer Cr$_2$S$_2$ consists of two Cr atoms and two S atoms. The optimized lattice constants for the Cr$_2$S$_2$ monolayer is 3.783 Å. The magnetic moments of the two Cr atoms are 3.404 $\mu_B$ and -3.404 $\mu_B$, respectively, resulting in a total magnetic moment of 0.00 $\mu_B$. The monolayer Cr$_2$S$_2$ possesses a square lattice structure, and its symmetry is described by the space symmetry group $P4-mmm$, which is a simple space group of the point group $D_{4h}$. The $D_{4h}$ point group contains $C_4^Z$, $C_2^X$ and $I$. The Brillouin zone for the band structure calculation is shown in  \hyperref[fgr1]{Fig. \ref{fgr1}(c)}, which is related to the crystal structure. To illustrate the stability of Cr$_2$S$_2$, phonon dispersion calculations are performed. By calculating the phonon dispersion of the 2×2×1 supercell, the dynamic stability of the structure is confirmed, as illustrated in \hyperref[fgr1]{Fig. \ref{fgr1}(d)}. Notably, no virtual frequencies are observed in the phonon spectrum, indicating that Cr$_2$S$_2$ is dynamically stable.
As mentioned previously, altermagnetism is a combination of AFM and FM. We constructed FM and AFM structures using 2×2×1 supercells to determine the antiferromagnetic properties of the Cr$_2$S$_2$ monolayer. As shown in \hyperref[fgr2]{Fig. \ref{fgr2}}, we investigat one FM and three AFM structures. The energies of FM and N\'{e}el AFM, zigzag AFM and stripy AFM configurations are listed in {Table \ref{tab:table1}}. It is proved that the energy of N\'{e}el AFM per unit cell is 960.3 meV, 283.5 meV and 370.5 meV lower than those of FM, zigzag AFM and stripy AFM cases. This confirms the N\'{e}el AFM configuration for the Cr$_2$S$_2$ monolayer.

\begin{table}[t]
\centering
\caption{\label{tab:table1}%
The energy difference between N\'{e}el AFM and FM, zigzag AFM and stripy AFM configurations, where AFM1 represents N\'{e}el AFM, AFM2 represents zigzag AFM, and AFM3 represents stripy AFM
 }
\begin{ruledtabular}
\begin{tabular}{ccc}
 $E_{FM-AFM1}$&$E_{AFM2-AFM1}$&$E_{AFM3-AFM1}$\\ 
 \hline
 960.3 meV&283.5.3 meV&370.5 meV \\
\end{tabular}
\end{ruledtabular}
\end{table}
\subsection{\label{sec:citeref}ELECTRONIC STRUCTURES}
\begin{figure}[t]
	\centering
	\includegraphics[width=\linewidth]{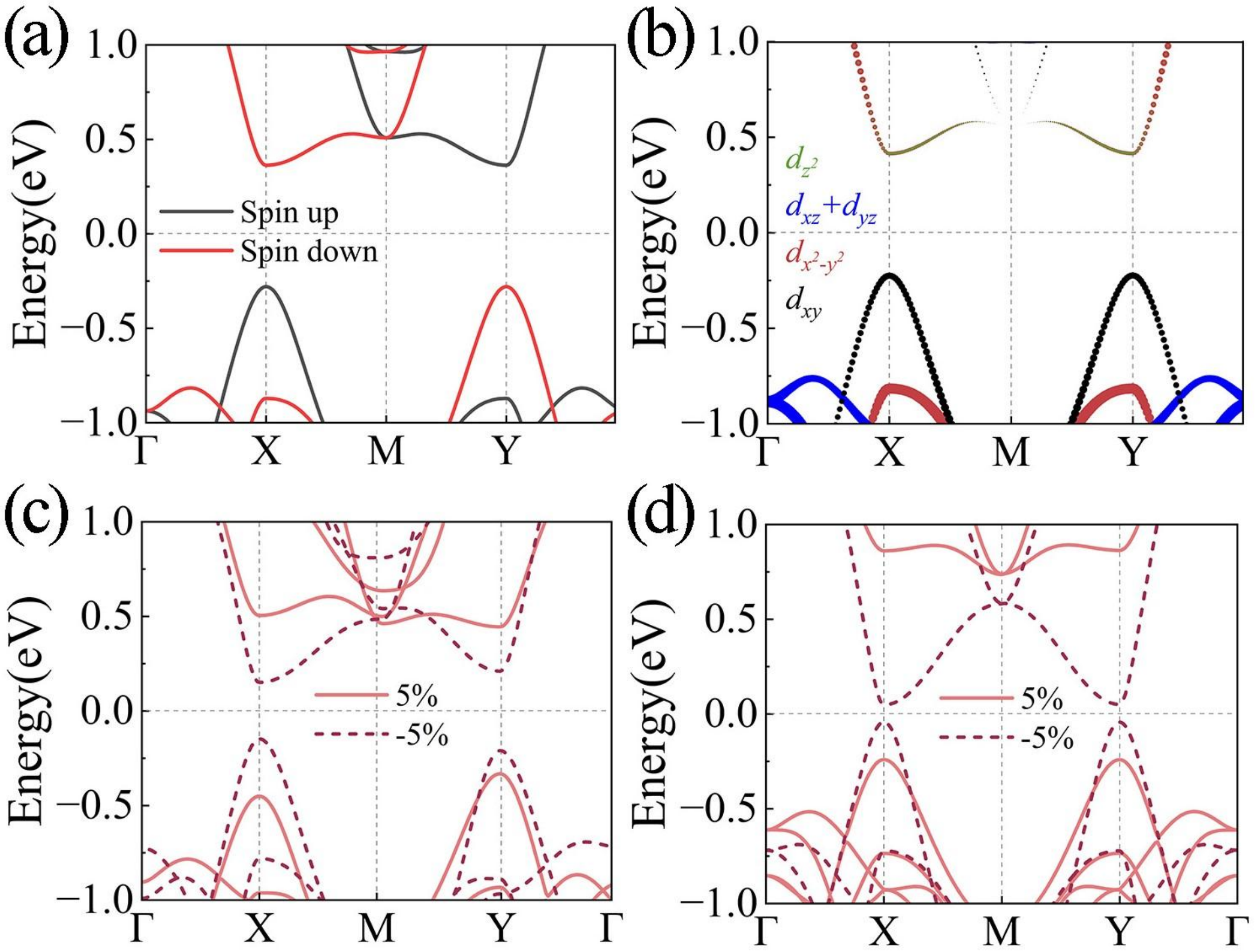}
	\caption{ Band structures of monolayer Cr$_2$S$_2$. (a) Band structures without considering SOC. (b) The projected PBE band structures for the Cr$_2$S$_2$ monolayers. Band structure evolution under different (c) strains along x direction (-5\%, 5\%) and (d) biaxial strain (-5\%, 5\%).
	}
	\label{fgr3}
\end{figure}

After confirming the AFM ground state of the Cr$_2$S$_2$ monolayer, we further investigate the band structure of the Cr$_2$S$_2$ monolayer. The band structure of the Cr$_2$S$_2$ monolayer without considering SOC is shown in \hyperref[fgr3]{Fig. \ref{fgr3}(a)}. There are two valleys at the X and Y points, and the states near these points mainly originate from two different Cr atoms. Unlike the general band structure of AFM\cite{RN48}, the band structure of the Cr$_2$S$_2$ shows spin$-$splitting phenomenon. In Cr$_2$S$_2$, the two sublattices Cr1$-$S and Cr2$-$S have different orientations, one in the x direction and the other in the y direction in \hyperref[fgr1]{Fig. \ref{fgr1}(a)}. The two sublattices are symmetrically $M_{xy}$ related by diagonal mirrors, which produces spin splitting. For the unstrained case, the gap between the X and Y valleys is the same, indicating that there is no valley polarization. Compared to Cr$_2$O$_2$\cite{RN722} without considering SOC, it can be clearly observed that there is no longer Weyl cones in the Cr$_2$S$_2$ band structure. This is because the onsite energy of S and O are different. Moreover, Cr$_2$S$_2$ is a direct band gap semiconductor with a gap value of 0.639 eV. The conductor band minimum (CBM) and the valence band maximum (VBM) of Cr$_2$S$_2$ come from bands with different spin directions. The Cr atom is affected by a crystal field with $D_{4h}$ symmetry, and the original five-fold degeneracy of the 3d orbitals is lifted and split into four groups: a non-degenerate A1 ($d_{z^2}$) orbital, two non-degenerate B1 ($d_{x^2-y^2}$) and B2 ($d_{xy}$) orbitals, and a doubly degenerate E ($d_{xz}$, $d_{yz}$) orbital. As shown in \hyperref[fgr3]{Fig. \ref{fgr3}(b)}, the VBM mainly comes from the $d_{xy}$ orbitals, while the CBM comes from the $d_{z^2}$ and $d_{x^2-y^2}$ orbitals.

To obtain valley polarization in Cr$_2$S$_2$, it is necessary to break the diagonal mirror symmetry $M_{xy}$. Strain engineering has been widely used to modulate the band structure of two-dimensional materials\cite{RN229}, and the symmetry of two-dimensional materials can be changed by applying uniaxial strain. Therefore, we apply uniaxial strain to adjust the band structure, which breaks the $M_{xy}$ symmetry so that the rotational symmetry $C_4$ is reduced to $C_2$. Strain ($\varepsilon$) is defined as $(a$-$a_0)/a_0$. Here, a and a$_0$ are lattice constants in the presence and absence of strain. \hyperref[fgr3]{Fig. \ref{fgr3}(c)} shows the band structure of a Cr$_2$S$_2$ monolayer considering uniaxial compressive and tensile strains. Compared to the unstrained case, the X, Y valleys still exist, but the energies at these valleys are not equal, breaking their degeneracy. For a strain of 5\%,  the band gap of the X valley energy decreases, while the band gap of the Y valley energy increases, resulting in a large valley polarization of the Cr$_2$S$_2$ monolayer. Valley polarization is defined as $\bigtriangleup E_{V(C)} = E_{V(C)}^Y - E_{V(C)}^X$. The valley polarization generated by the valence band, denoted as $\bigtriangleup E_{V} = -118.39~meV$ and the valley polarization $\bigtriangleup E_{C} = 59.81~meV$ in the conduction band, sum up to a total valley polarization of $\bigtriangleup E_{VCP} = 178.20~meV$ ($\bigtriangleup E_{VCP} = \bigtriangleup E_{C} $-$ \bigtriangleup E_{V}$). The valley splitting of the Cr$_2$S$_2$ is higher than the valley splittings of many widely studied ferrovalley materials.\cite{RN575,RN635,RN229} 

Compared with 5\%, the valley polarization reverses when the strain is $-$5\%. Meanwhile, the strain in the opposite direction produces the exact opposite valley polarization, the VBM and the CBM are both in the Y valley. From compressive to tensile strain, the band band structure further away from the Fermi level. The coupling property between valley polarization and strain direction allows us to use strain control valley and apply it to logic devices, which can be applied to logic devices.In addition, we also exhibit the change of band structure under biaxial strain. As delivered \hyperref[fgr3]{Fig. \ref{fgr3}(d)}, Under 5\% and -5\% biaxial strain, the energies of the X and Y valleys are equal, resulting in no valley polarization. This is because biaxial strain does not break the diagonal mirror symmetry $M_{xy}$.

\begin{figure}[t]
	\centering
	\includegraphics[width=\linewidth]{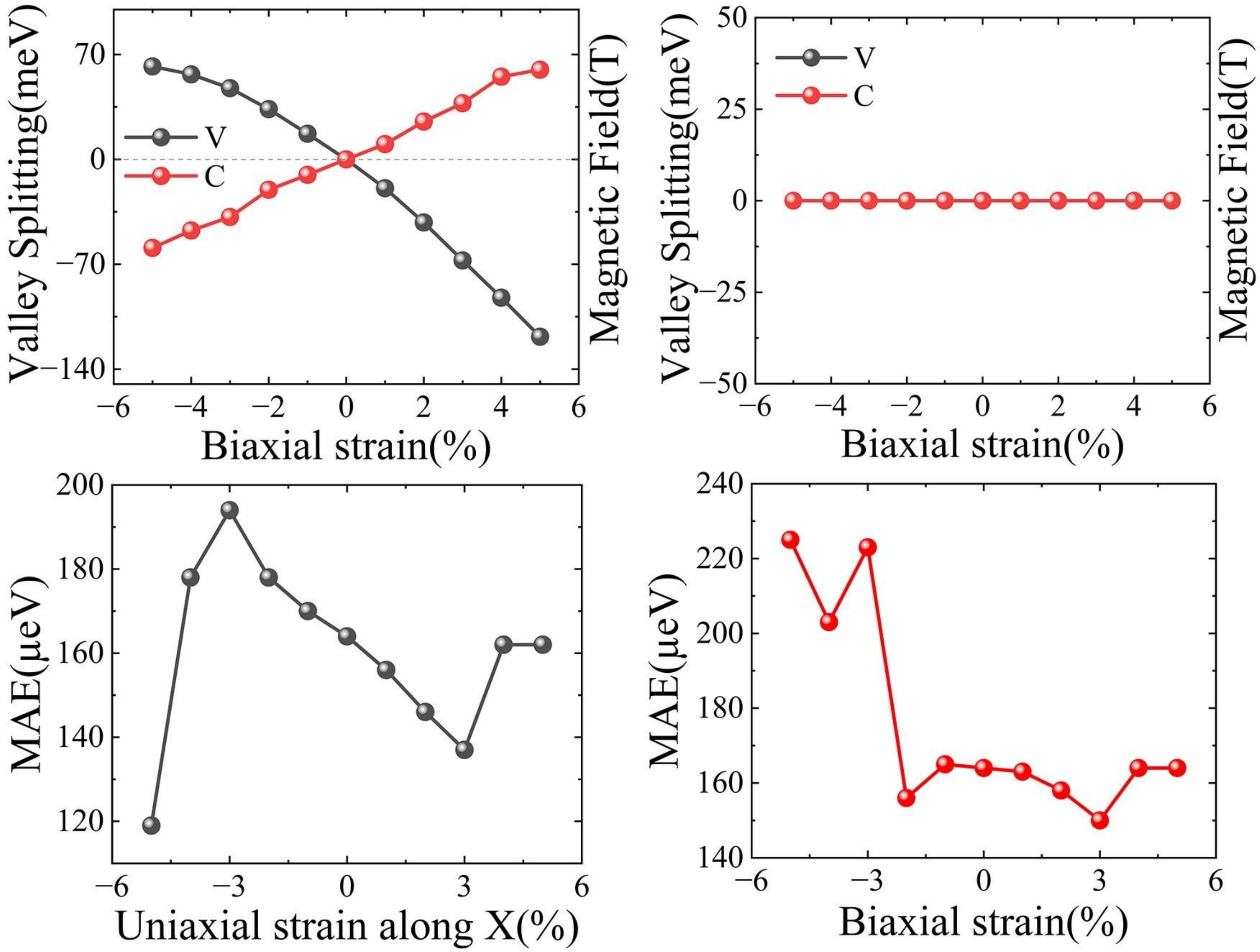}
	\caption{Variations in band structure, effective magnetic field, and MAE under strain. The changes in valley polarization of the valence band and conduction band of the Cr$_2$S$_2$ monolayer are presented for (a) uniaxial strain along the X direction and (b) biaxial strain, along with the corresponding variations in the effective magnetic field. MAE under different (c) uniaxial strains along X and (d) biaxial strain. The strain range is $-$5\% to 5\%. 
	}
	\label{fgr4}
\end{figure}
\begin{figure*}[t!]
\centering
\includegraphics[width=0.85\textwidth]{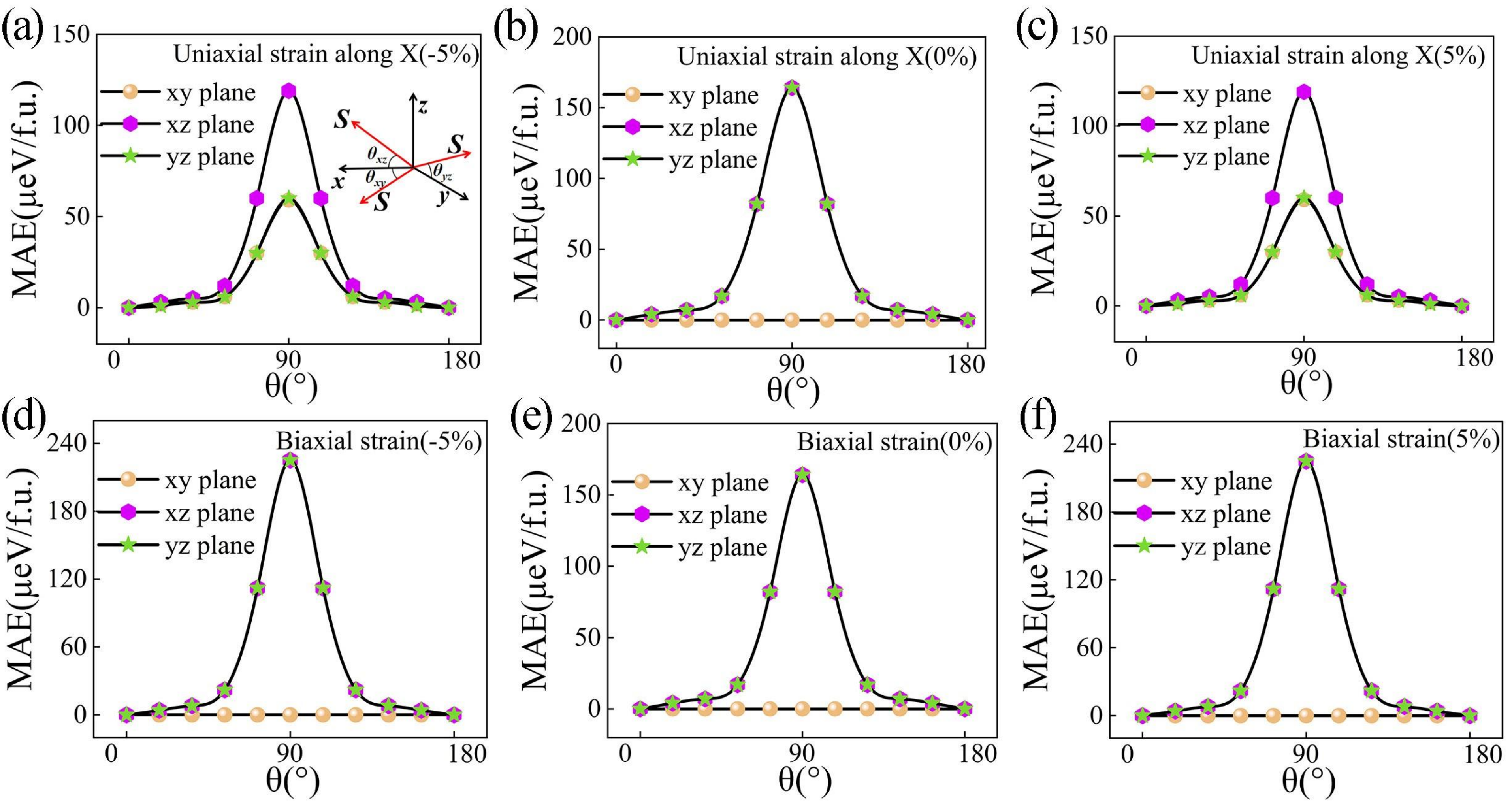}
\caption{\label{fig:wide}MAE of the Cr$_2$S$_2$ monolayers along xy, xz and yz planes. Under different (a$-$c) uniaxial strains along x direction ( $- 5\%$, 0\%, and 5\%) and (d$-$f) biaxial strains ($- 5\%$, 0\%, and 5\%).}
\label{fgr5}
\end{figure*}

The evolution of valley splitting under strain are plotted in \hyperref[fgr3]{Fig. \ref{fgr4}(a) and (b)}. Consistent with the above analysis, uniaxial strain can induce valley polarization. And with the increase of compressive and tensile strength, valley splitting also increases, which is conducive to the experimental manipulation of these valleys. In the case of biaxial strain, no matter how we regulate the strain, there will be no valley polarization. The specific changes in the band structure under strain are shown in Figures S2 and S3. In summary, symmetry breaking is the key to valley polarization of altermagnet.

It can also be observed from \hyperref[fgr3]{Fig. \ref{fgr3}(c)} that the conduction band (valence band) around the X (Y) valley has almost the same energy displacement under the action of uniaxial strain, indicating that the uniaxial strain induces a static and uniform effective magnetic field. The strength of the induced effective magnetic field can be expressed as\cite{RN401}
\begin{equation}\label{eqn-1}
B_{c(v)}^{\mathrm{eff}} \equiv \frac{\delta E^{V / C}}{g_s s_z},
\end{equation}
where $\delta \mathrm{E}^{\mathrm{V} / \mathrm{C}}$ is the size of valley splitting of conduction and valence band, $g_s=1/2$, $s_z=2$. Equation\eqref{eqn-1} can be expressed as follows:
\begin{equation}\label{eqn-13}
B_{c(v)}^{e f f}=\delta E^{V / C}=\eta^{V / C} \varepsilon_{x x}. 
\end{equation}
The variation of the effective magnetic field is consistent with Equation\eqref{eqn-7}. As shown in \hyperref[fgr4]{Fig. \ref{fgr4}(a)}, the effective magnetic field $B_{c(v)}^{e f f}$ and the valley splitting $\delta{E}$ exhibit the same trend under uniaxial strain. With a 5\% uniaxial strain, effective magnetic fields $B_{v}^{\mathrm{eff}}=118.39T$ and $B_{c}^{\mathrm{eff}}=59.81T$ are produced. This indicates that the altermagnetic materials can generate an effective magnetic field under small strains.

Magnetic anisotropy energy(MAE) is one of the key factors to realize long$-$range magnetic order in two-dimensional materials, which is directly related to the thermal stability of magnetic data storage. MAE is represented as $MAE=E[001]$-$E[100]$, where E[100] and E[001] are the system energies when the magnetic moment is along the in$-$plane [100] axis and the out-of-plane [001], respectively. In this study, we used $GGA $+$ U $+$ SOC$ method to calculate the total energy in the direction of magnetic moment [100], [001]. The MAE value of the Cr$_2$S$_2$ monolayer is $-$0.164meV, with $E[100]$-$E[100]<0$, indicating in$-$plane magnetic anisotropy (IMA) character. \hyperref[fgr4]{Fig. \ref{fgr4}(c) and (d)} demonstrates the effects of uniaxial and biaxial strains on MAE. The Cr$_2$S$_2$ monolayer maintains IMA behavior throughout the strain range, this shows the robustness of MAE under strain.

To further analyze the characteristics of MAE in the whole space, we calculate the projection of MAE on the xy, xz and yz planes, as observed in \hyperref[fgr5]{Fig. \ref{fgr5}}. It is clear that MAE exhibits a strong dependence on the direction of magnetization along the xz and yz planes, independent of the xy planes in \hyperref[fgr5]{Fig. \ref{fgr5}(b)}.and \hyperref[fgr5]{Fig. \ref{fgr5}(e)}. When uniaxial strain is applied, it changes the rotational symmetry from $C_4$ to $C_2$, leading to symmetry breaking. At this point, the MAE is anisotropic in the xy, xz, and yz planes. This phenomenon does not occur under biaxial strain. The MAE remains isotropic in the xy plane and anisotropic in the xz and yz planes. This is because biaxial strain does not change the symmetry, and the rotational symmetry remains as $C_4$. From Figure S4 and Figure S5, it can be seen that
the contribution of the Cr-p orbitals is relatively small, an order of magnitude less than that of the d orbitals. As the compressive strain increases, the contribution of $d_{z^2}$ and $d_{yz}$ hybridization to the anisotropy in the [010] direction decreases, while the contribution of $d_{z^2}$ and $d_{xy}$ hybridization to the anisotropy in the [100] direction also decreases. Overall, under compressive strain, the reduction in the contribution of Cr's d orbitals along the [100] direction is greater than that along the [010] direction. This disparity results in the emergence of anisotropy in the xy plane under uniaxial strain conditions.

\subsection{\label{sec:citeref}PIEZOMAGNETISM IN MONOLAYER}
As mentioned above, Cr$_2$S$_2$ monolayer can easily generate valley polarization using uniaxial strain, which provides new opportunities to generate net magnetization. Piezomagnetic is generated by the inherent magnetoelastic coupling, which is a static property. This coupling can also lead to magnetostriction where one can induce a strain using an external magnetic field. Strain-induced magnetization is obtained as\cite{RN686}:
\begin{equation}\label{eqn-9}
	\begin{aligned}
M\left(\theta, \varepsilon_\theta, n\right)= 
\begin{cases}-\gamma_{\phi\theta} \varepsilon_\theta, & 0 \leq \varepsilon_\theta<\frac{n}{|\gamma_{\phi\theta}|} \\ -\operatorname{sign}(\gamma_{\phi\theta}) n, & \varepsilon_\theta \geq \frac{n}{|\gamma_{\phi\theta}|}
\end{cases}
	\end{aligned},
\end{equation}
where $\gamma_{\phi\theta}=\rho\left[\left(D_{x x}-D_{y y}\right) \sin 2 \phi-2 D_{x y} \cos 2 \phi\right]\sin (2 \phi-2 \theta)=\gamma\sin (2 \phi-2 \theta)$, $\gamma$ is a constant, $\phi$ is the angle between the mirror and the X$-$axis, $\theta$ is the angle between the direction of uniaxial strain applied and the X$-$axis. In this paper, $\phi=\pi/4$, $\theta=0$. Uniaxial strain-induced magnetization is obtained as can be written as
\begin{equation}\label{eqn-10}
	\begin{aligned}
M= 
\begin{cases}-\gamma \varepsilon, &  0 \leq \varepsilon<\frac{n}{|\gamma|}\\ -\operatorname{sign}(\gamma) n, & \varepsilon \geq \frac{n}{|\gamma|}
\end{cases}
	\end{aligned},
\end{equation}
according to the Equation\eqref{eqn-10}, the relationship between magnetization and strain is linear under finite strain. With the increase of strain, the magnetization becomes saturated. Similarly, the biaxial strain magnetisation M is 0. 
\begin{figure}[t]
	\centering
	\includegraphics[width=\linewidth]{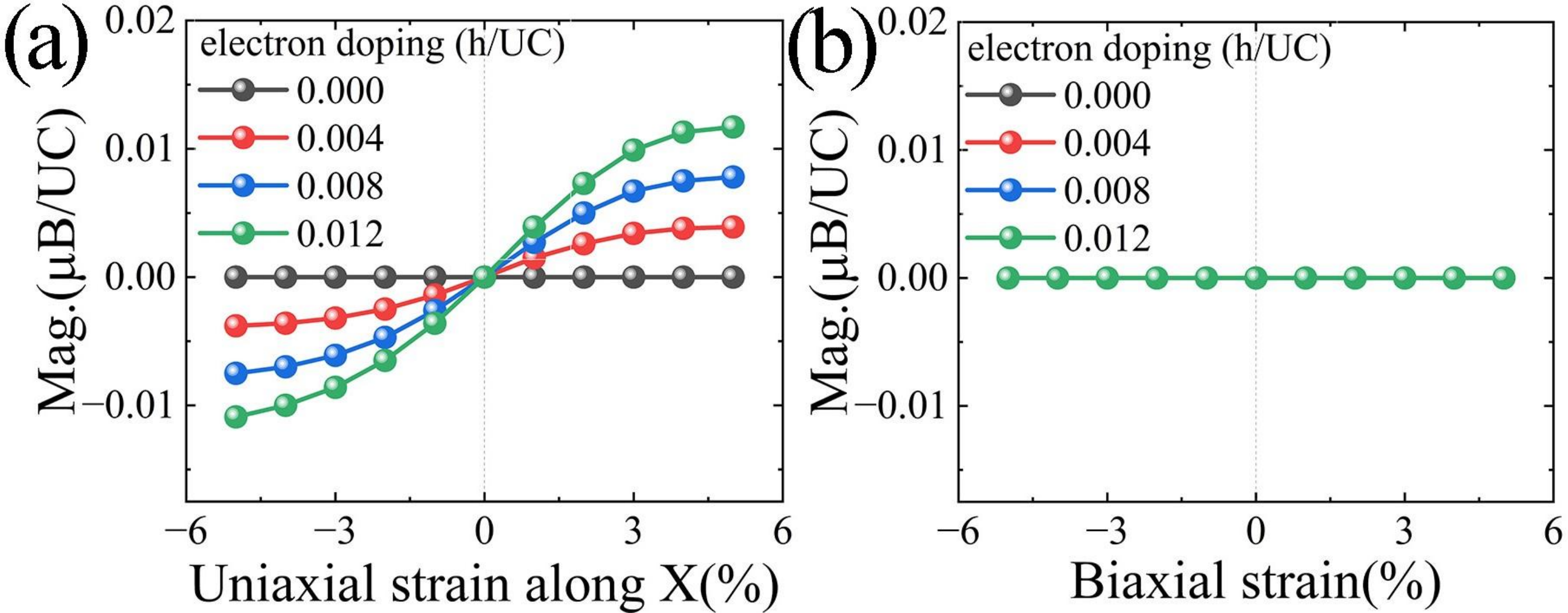}
	\caption{Strain$-$induced piezomagnetic in electron$-$doped monolayer Cr$_2$S$_2$. The net magnetization change under (a) uniaxial strains along X and (b) biaxial strains.
}
	\label{fgr6}
\end{figure}
To accurately determine the magnetization during electron doping, we tested the K-point convergence, as illustrated in Figure S6. \hyperref[fgr6]{Fig. \ref{fgr6}(a)} displays the magnetization in relation to carriers density and uniaxial strain. The magnetization increases with both carrier density and strain, and the direction of magnetization is indeed reversed for tensile and compressive strains. In the absence of strain, there is no net magnetic moment because. This is because the diagonal mirror symmetry $M_{xy}$ is preserved. When the strain is small, the magnetization increases linearly with the strain, which is consistent with the description of Equation\eqref{eqn-10}. As the strain strength increases gradually, the magnetization tends to be saturated. This is consistent with our previous analysis. One can see from \hyperref[fgr6]{Fig. \ref{fgr6}(b)} that since the symmetry is not broken, the biaxial strain does not change the magnetization and remains zero. Compared with the piezomagnetism in nonlinear antiferromagnetic materials\cite{RN785}, the piezomagnetism of Cr$_2$S$_2$ is much larger, and there are ways to adjust the size and direction of magnetization. Given a strain intensity, the magnetization magnitude can be adjusted by the electron doping concentration. Given the electron doping concentration, the direction of magnetization can be changed by uniaxial strain. The piezomagnetic properties of Cr$_2$S$_2$ make it a good candidate for magnetic sensors and other applications.

\section{CONCLUSION}
In summary, we investigate the influence of external field modulation on valley polarization. Utilizing DFT calculations, we systematically examine several properties of the Cr$_2$S$_2$ monolayer, including its stability, magnetic ground state, band structure, and magnetic anisotropy. The Cr$_2$S$_2$ monolayer demonstrates excellent stability with an AFM magnetic ground state and intriguing spin valley splitting, establishing it as an altermagnetic material. Valley polarization can be manipulated through uniaxial strain, with the direction of polarization altered by varying between compressive and tensile strains. However, such polarization control is absent under biaxial strains, which maintain the diagonal mirror symmetry $M_{xy}$ and preserve the $C_4$ rotational symmetry. We find that an almost uniform effective magnetic field is generated under uniaxial strain, which provides favorable conditions for applications in spintronic strain control. The MAE of Cr$_2$S$_2$ monolayers remains robust and in-plane under both uniaxial and biaxial strains. Further analysis of MAE in the whole space shows that MAE is isotropic in the xy plane under the absence of strain and biaxial strain. However, under uniaxial strain, MAE exhibits anisotropy in xy, zx and yz planes. Remarkably, under the influence of uniaxial strain, the Cr$_2$S$_2$ monolayer exhibits piezomagnetic characteristics. Modulating hole doping and adjusting strain strength allow precise control over the size and orientation of magnetization. Our findings contribute to the advancement of altermagnetic and valleytronic materials, promising significant potential for energy$-$efficient and ultra$-$fast valleytronic devices.

	We acknowledge the fundings from National Natural Science Foundation of China (Grant No. 51872145), Postgraduate Research \& Practice Innovation Program of Jiangsu Province (Grant No. KYCX23\_0977), and Qinglan Project of Jiangsu Province of China.

\bibliography{apssamp}

\end{document}